\documentclass[showpacs,preprint]{revtex4}
\usepackage{amsfonts}
\usepackage{amsmath}
\usepackage{amssymb}
\usepackage{hyperref}

\begin{document}

\title{General stationary charged black holes as charged particle
accelerators}
\author{Yi Zhu$^{1}$}
\author{Shao-Feng Wu\footnote{%
Corresponding author. Email: sfwu@shu.edu.cn; Phone: +86-021-66136202.}$%
^{1,2}$}
\author{Yu-Xiao Liu$^{3}$}
\author{Ying Jiang$^{1,2}$}
\affiliation{$^{1}$Department of physics, Shanghai University, Shanghai, 200444, P. R.
China}
\affiliation{$^{2}$The Shanghai Key Lab of Astrophysics, Shanghai, 200234, P. R. China}
\affiliation{$^{3}$Institute of Theoretical Physics, Lanzhou University, Lanzhou, 730000,
P. R. China}

\begin{abstract}
We study the possibility of getting infinite energy in the center of mass
frame of colliding charged particles in a general stationary charged black
hole. For black holes with two-fold degenerate horizon, it is found that
arbitrary high center-of-mass energy can be attained, provided that one of
the particle has critical angular momentum or critical charge, and the
remained parameters of particles and black holes satisfy certain
restriction. For black holes with multiple-fold degenerate event horizons,
the restriction is released. For non-degenerate black holes, the ultra-high
center-of-mass is possible to be reached by invoking the multiple scattering
mechanism. We obtain a condition for the existence of innermost stable
circular orbit with critical angular momentum or charge on any-fold
degenerate horizons, which is essential to get ultra-high center-of-mass
energy without fine-tuning problem. We also discuss the proper time spending
by the particle to reach the horizon and the duality between frame dragging
effect and electromagnetic interaction. Some of these general results are
applied to braneworld small black holes.
\end{abstract}

\maketitle
\affiliation{Institute of Theoretical Physics, Shanghai University}

\section{Introduction}

Recently, Ba\~{n}ados, Silk and West (BSW) \cite{BSW Mechanic} proposed a
mechanism to obtain unlimited center-of-mass (CM) energy of two particles
colliding at an extreme Kerr black hole (BH), which was henceforth asserted
as a natural Planck-scale particle accelerator and the possible origins for
the very highly energetic astrophysical phenomena, such as the gamma ray
bursts and the active galactic nuclei. However, authors of \cite{Comment}
and \cite{Jacobson} pointed out that the collision in fact takes an infinite
proper time. Moreover, even the ultra-energetic collisions cannot occur in
nature, because the astrophysically limited maximal spin prohibits the
formation of real extreme Kerr BHs, meanwhile the back-reaction effects may
reduce the allowed maximized CM energy below the Planck scale upon
absorption of a first pair of colliding particles, and the gravitational
radiation of the particle with fine-tuning critical angular momentum should
be peaked at frequencies corresponding to marginally bound quasi-circular
orbits.

To achieve arbitrary high CM energy under the limitation of maximal spin,
the multiple scattering mechanism was taken into account in the nonextremal
Kerr BH \cite{pavlov}. Another more direct method is to consider different
extreme rotating BHs, such as Kerr-Newman BHs \cite{KN}. Actually, by making
use of a special metric which is convenient to discuss the near-horizon
geometry of general axially symmetric rotating BHs, Zaslavskii showed the
unbound CM energy of colliding particles at the extreme horizon or
nonextremal horizon by considering multiple scattering \cite{general
rotating}. Bearing in mind the universal character, it was pointed out that
the potential acceleration to large energies should be taken seriously both
as manifestation of general properties of BHs and the effect relevant in
astrophysics of high energies. Other concrete models were also studied and
some new results which have not been included in the general frame appeared,
such as the Kerr-(A)dS BH where three-fold degenerate horizons were
considered \cite{Jie Yang}. In addition, it was pointed out that the
existence of innermost stable circular orbit (ISCO) in Kerr BHs would avoid
the artificial fine-tuning in an astrophysical context \cite{ISCO}. See more
backgrounds like Sen BHs \cite{Sen}, Kaluza-Klein BHs \cite{KK},
Kerr-Taub-NUT BHs \cite{Chen}, and naked singularities \cite{naked
singularity}. Noticing these alternative options for generating extremal
black holes, the BSW mechanism has been further studied by calculating the
escaping flux of massless particles for maximally rotating black holes, and
it was suggested that the received spectrum should typically contain
signatures of highly energetic products \cite{Banados}, see also the sequent
numerical estimation \cite{Williams}.

In the aforementioned mechanisms for ultra-high CM energy, the frame
dragging effect of rotating BHs is necessary. However, Zaslavskii \cite{RN}
showed that a non-rotating but charged Reissner-Nordstr\"{o}m (RN) BH can
also serve as an accelerator with arbitrarily high CM energy of charged
particles collided at the extreme horizon or nonextremal horizon considering
multiple scattering. It was demonstrated that the upper bound of the
electric charge of BHs after Schwinger emission is large enough to allow the
ultra-high CM energy of charged colliding particles. The only restriction is
that the BH should not be too light ($>10^{20}g$). Moreover, noticing the
correspondence between frame dragging effect and the electromagnetic
interaction as well as the higher symmetry of RN BHs than Kerr BHs, an upper
bound was suggested to exist for the total energy of colliding particles in
the observable domain in the BSW process due to the gravity of the particles
\cite{shell}.

In this paper, we will investigate the BSW mechanism with the combined
effect of frame dragging and electromagnetic interaction. Instead of
restricting on a concrete charged BH, we will study a general background
with charged test particles. Instead of the special metric adopted in \cite%
{general rotating}, we will use a general stationary metric which can be
reduced to Kerr metric using Boyer-Lindquist coordinates directly and hence
can be related to observations (such as escape fraction) conveniently. We
will not only discuss the universal existence of ultra-energetic collisions
at usual two-fold degenerate horizons or nonextremal horizons considering
multiple scattering, the associated proper time and the fine-tuning critical
angular momentum or charge, but also discuss multiple-fold degenerate
horizons and ISCO in general. Some new restrictions on parameters of
particles and BHs, which are necessary for ultra-energetic collisions but
have not been noticed before, will be revealed as well. Moreover, we will
show the correspondence between frame dragging and electromagnetic
interaction in BSW mechanism in more detail.

On the other hand, based on the well-known braneworld scenario with large or
compact extra dimensions, the fundamental Planck scale is lowered to the
order of magnitude around the TeV scale. A particularly exciting proposal is
the possibility of creating mini BHs in super colliders, such as LHC, which
can achieve the energy scale about 14 TeV and could be taken as the
\textquotedblleft factory\textquotedblright\ of small BHs \cite{Kanti}. This
provides a potential terrestrial check of phenomenon around astrophysical
BHs, including the BSW mechanism. As an application and in fact a strong
motivation of our general frame, we will discuss the collision of charged
test particles in the Arkani-Hamed, Dimopoulos and Dvali (ADD) braneworld
Kerr-Newman (KNM) BH, which was suggested recently in \cite{Sampaio}. As it
was pointed out, due to the strengthening of the gravitational field
compared to the electromagnetic field in TeV gravity scenarios, the initial
evaporation is not dominated by fast Schwinger discharge. So, the braneworld
KNM BH is a general BH which could be formed after proton-proton collisions
in LHC, providing the possible further signatures of BH events such as
charge asymmetries. We will show that there is no degenerate horizon, but
the ultra-energetic collisions can be produced by considering multiple
scattering. To be complementary, we will discuss the BSW process near the
degenerate horizon of the tidal charged BH based on Randall-Sundrum (RS)
braneworld scenarios, which is similar to the 4-dimensional KNM BH, but the
tidal charge could be large ($<10^{4}m^{2}$ in solar system tests) \cite%
{hmer}.

The rest of paper is arranged as follows. In section II, we will give the
general gravitational background, the geodesic motion of charged particles
under electromagnetic field, and the CM energy of colliding particles. In
section III, we will analyze the effective potential to determine whether
the particle can reach the horizon, and the CM energy will be studied for
the nonextremal horizon considering the multiple scattering, the two-fold
degenerate horizon, and the multiple-fold degenerate horizon, respectively.
The proper time, possible ISCO, and the duality between frame dragging and
coulomb interaction in BSW mechanism will also be investigated. In section
IV, we will apply some obtained results in the general frame to the
braneworld BH background. The final section is devoted to conclusion and
discussion.

\section{Geodesic motion of two charged particles}

Since the BSW mechanism relates to the geodesic motion of test particles in
the gravitational background, we would like to analyze equations of their
motion in detail. For simplification, we assume that the motion of particles
occurs in the equatorial plane ($\theta =\pi /2$) of BHs (for the
non-equatorial motion in the Kerr BH see \cite{Harada}). For a general BH,
the metric can be written as
\begin{equation}
ds^{2}=\left[ -g_{1}(r)+g_{2}(r)w^{2}(r)\right] dt^{2}-2g_{2}(r)w(r)dtd\phi +%
\frac{dr^{2}}{g_{3}(r)}+g_{2}(r)d\phi ^{2},  \label{metric}
\end{equation}%
where $g_{1}(r)$, $g_{2}(r)$, $g_{3}(r)$, and $w(r)$ are arbitrary functions
of the radial coordinate $r$. Compared with the metric adopted in \cite%
{general rotating} where $g_{3}(r)=1$ and $g_{1}(r)=N^{2}$ (with the horizon
located at $N=0$), the metric (\ref{metric}) is more convenient to compare
with Kerr and RN cases (with the horizon located at $g_{3}(r)=0$) and can be
directly imposed with asymptotically flat which is important to compare with
observations. One may worry about the complicated near-horizon geometry with
metric (\ref{metric}) which was simplified in the metric adopted in \cite%
{general rotating}, but we will show later that it still can be tackled when
noticing the near-horizon behavior of $g_{1}(r)$ and $g_{3}(r)$. The
electromagnetic potential of BHs are denoted as $A_{t}=B(r)$, $A_{\phi
}=C(r) $, which are treated as arbitrary functions of $r$. Other components
of the electromagnetic potential vanish due to the symmetry of background.
Because the functions of metric are independent to the coordinates $t$ and $%
\phi $, there are two Killing vectors for the BH, expressed as $\xi ^{\mu
}=(1,0,0,0) $ and $\eta ^{\mu }=(0,0,0,1)$. For a test particle with rest
mass $m_{0}$ and charge $q$ per unit rest mass, it should has two conserved
quantities
\[
-\xi ^{\mu }\left( u_{\mu }-qA_{\mu }\right) =E\quad \text{and}\quad \eta
^{\mu }\left( u_{\mu }-qA_{\mu }\right) =L,
\]%
where $E$ and $L$ are the conserved parameters of the energy and angular
momentum of the test particle per unit rest mass, respectively. By combining
these equations with the timelike restriction of test particles $u\cdot u=-1$%
, the geodesic equations can easily be solved as%
\begin{eqnarray}
u^{t} &=&\frac{\Xi }{g_{1}},  \label{ut} \\
u^{r} &=&-\sqrt{\frac{g_{3}}{g_{1}g_{2}}\left[ g_{2}\Xi ^{2}-g_{1}\left(
g_{2}+\Lambda ^{2}\right) \right] },  \label{ur} \\
u^{\phi } &=&\frac{\Lambda g_{1}+g_{2}w\Xi }{g_{1}g_{2}},
\end{eqnarray}%
where
\[
\Lambda _{i}(r)=L_{i}+q_{i}C(r),\;\Theta _{i}(r)=E_{i}-q_{i}B(r),\;\text{%
and\quad }\Xi _{i}(r)=\Theta _{i}(r)-\Lambda _{i}(r)w(r),
\]%
and index $i=1$, $2$ denotes two different particles. Note that the minus
sign in Eq. (\ref{ur}) means ingoing particles that we are concerning, for
the discussion on outgoing particles please see \cite{general explanation}.

From Eq. (\ref{ur}), we find that there is a particular interesting case in
which $g_{1}(r)$ and $g_{3}(r)$ have to tend to zero with the same speed,
otherwise the particle will have a vanishing or divergent radial velocity on
the horizon. This case is also required to be consistent with the static
vacuum background where $g_{1}(r)=g_{3}(r)$. Moreover, considering that the
three-fold degenerate horizons might take role in BSW mechanism \cite{Jie
Yang}, we will try to study what will happen near general $n$-fold
degenerate horizons (For convenience later, we will call $n=1$ as
non-degenerate horizons, $n\geq 2$ as any-fold degenerate horizons, and $%
n\geq 3$ as multiple-fold degenerate horizons.) For this aim, we take the
replacement
\begin{equation}
g_{1}(r)\rightarrow (r-r_{H})^{n}g_{6}(r),\qquad g_{3}(r)\rightarrow
(r-r_{H})^{n}g_{5}(r),  \label{replacement}
\end{equation}%
where $r_{H}$ denotes the location of horizon. Here we would like to stress
that this replacement is a simple but key step to discuss the behavior near
any-fold degenerate horizons. Furthermore, we note that, for the region
outside the event horizon (while not the Cauchy horizon, see \cite{Lake}), $%
g_{5}(r_{H})$ and $g_{2}(r_{H})$ are positive definite to preserve $r$ and $%
\phi $ as spatial coordinates. $g_{6}(r_{H})$ should also be positive
definite, or there is no observer at fixed $r$ and $\theta $ will be
permitted.

Now, let us discuss the CM energy of two particles with the same rest mass $%
m_{0}$ colliding in this background by calculating%
\begin{equation}
E_{c.m.}=\sqrt{2}m_{0}\sqrt{1-u_{(1)}\cdot u_{(2)}}.  \label{Ecm}
\end{equation}%
The result is%
\begin{equation}
\frac{E_{c.m.}^{2}}{2m_{0}^{2}}=1+\frac{1}{g_{1}g_{2}}\left[ g_{2}\Xi
_{1}\Xi _{2}-g_{1}\Lambda _{1}\Lambda _{2}-\sqrt{g_{2}\Xi
_{1}^{2}-g_{1}(\Lambda _{1}^{2}+g_{2})}\sqrt{g_{2}\Xi _{2}^{2}-g_{1}(\Lambda
_{2}^{2}+g_{2})}\right] .  \label{uc0}
\end{equation}%
One can find that the numerator and denominator of the fraction in the above
equation are both vanishing on the horizon where $g_{1}(r)=0$, if $\Xi
_{i}\geqslant 0$ (We do not consider the case with $\Xi _{i}<0\,$, because
in the next section, we will show that the particles with $\Xi _{i}<0$ can
not fall into the horizon.). Thus, the fraction is underdetermined on the
horizon. To analyze this underdermination, one needs to consider the
near-horizon behavior by the replacement (\ref{replacement}). Hence, Eq. (%
\ref{uc0}) becomes to%
\begin{eqnarray}
\frac{E_{c.m.}^{2}}{2m_{0}^{2}} &=&1+\frac{1}{(r-r_{H})^{n}g_{2}g_{6}}\bigg[%
g_{2}\Xi _{1}\Xi _{2}-g_{6}\Lambda _{1}\Lambda _{2}(r-r_{H})^{n}  \nonumber
\\
&&-\sqrt{g_{2}\Xi _{1}^{2}-(r-r_{H})^{n}g_{6}(\Lambda _{1}^{2}+g_{2})}\sqrt{%
g_{2}\Xi _{2}^{2}-(r-r_{H})^{n}g_{6}(\Lambda _{2}^{2}+g_{2})}\bigg].
\label{uc1}
\end{eqnarray}

\section{BSW mechanism for general cases}

\subsection{Critical angular momentum and critical charge}

By making use of the l'Hospital's rule to do $n$th-order differential on the
numerator and denominator at the same time, the CM energy on the horizon can
be obtained from Eq. (\ref{uc1}). In the denominator, the terms remaining
with $r-r_{H}$ after the $n$th-order differential will vanish while $%
r\rightarrow r_{H}$, and the only term left is $\sim g_{2}(r)g_{6}(r)$, no
matter what the value of $n$ is. The situation for the numerator is similar,
as we will show below. One can rewrite the numerator of Eq. (\ref{uc1}) as
\begin{eqnarray*}
&&g_{2}\Xi _{1}\Xi _{2}-g_{6}\Lambda _{1}\Lambda _{2}(r-r_{H})^{n}- \\
&&\sqrt{\left( g_{2}\Xi _{1}\Xi _{2}\right) ^{2}-g_{2}g_{6}[\Xi
_{1}^{2}(\Lambda _{2}^{2}+g_{2})+\Xi _{2}^{2}(\Lambda
_{1}^{2}+g_{2})](r-r_{H})^{n}+g_{6}^{2}(\Lambda _{1}^{2}+g_{2})(\Lambda
_{2}^{2}+g_{2})(r-r_{H})^{2n}},
\end{eqnarray*}%
Respecting that $(r-r_{H})^{n}$ is a small quantity near the horizon, one
can expand the last term of above equation as following:%
\[
g_{2}\Xi _{1}\Xi _{2}-\frac{g_{2}g_{6}[\Xi _{1}^{2}(\Lambda
_{2}^{2}+g_{2})+\Xi _{2}^{2}(\Lambda _{1}^{2}+g_{2})](r-r_{H})^{n}}{%
2g_{2}\Xi _{1}\Xi _{2}}+\mathcal{O}(r-r_{H})^{2n}.
\]%
The numerator hence can be simplfied to%
\begin{equation}
\frac{(r-r_{H})^{n}g_{6}\left[ \left( \Lambda _{2}\Xi _{1}-\Lambda _{1}\Xi
_{2}\right) ^{2}+g_{2}(\Xi _{1}^{2}+\Xi _{2}^{2})\right] }{2\Xi _{1}\Xi _{2}}%
.  \label{rg}
\end{equation}%
Considering Eq. (\ref{rg}) as the form of a function multiplying $%
(r-r_{H})^{n}\,$ and using the Leibniz formula of high-order derivatives,
one can calculate the $n$-th order differential of Eq. (\ref{rg}) with
respect to $r$ near horizons:%
\[
n!\frac{g_{6}\left[ \left( \Lambda _{2}\Xi _{1}-\Lambda _{1}\Xi _{2}\right)
^{2}+g_{2}(\Xi _{1}^{2}+\Xi _{2}^{2})\right] }{2\Xi _{1}\Xi _{2}}%
|_{r\rightarrow r_{H}}.
\]%
Finally, for arbitrary integer $n$, the CM energy can be expressed as a very
simple form%
\begin{equation}
\frac{E_{c.m.}^{2}}{2m_{0}^{2}}=1+\frac{\left( \Lambda _{2}\Xi _{1}-\Lambda
_{1}\Xi _{2}\right) ^{2}+g_{2}(\Xi _{1}^{2}+\Xi _{2}^{2})}{2g_{2}\Xi _{1}\Xi
_{2}}|_{r\rightarrow r_{H}}.  \label{CMn}
\end{equation}%
It is intersting to see the indepdence with $n$.

The CM energy is infinite if one test particle has the critical angular
momentum or critical charge, which can be solved by setting $\Xi _{i}$ to
zero, given by:
\begin{equation}
L_{c}=\frac{E-qB-qCw}{w}|_{r\rightarrow r_{H}},  \label{Lc}
\end{equation}%
and%
\begin{equation}
q_{c}=\frac{E-Lw}{B+Cw}|_{r\rightarrow r_{H}}.  \label{qc}
\end{equation}%
These two equations indicate that the critical angular momentum and critical
charge of a test particle are entangled. In other word, for a test particle
with large angular momentum, the infinite CM energy can be attained for the
particle with small charge, and vice verse.

The critical angular momentum and charge can also be obtained by imposing $%
-\chi \cdot u=\Xi $ as zero at horizon. Here $\chi =\xi +w(r_{H})\eta $ is
the Killing vector generating the horizon. In \cite{Harada}, it was
conjectured that the CM energy is unbound if and only if the ratio $\frac{%
\chi _{2}\cdot p_{2}}{\chi _{1}\cdot p_{1}}$ is zero or infinite at Killing
horizon, where $p$ is the momentum. Now we find that the conjecture still
holds if $p$ is replaced with $m_{0}u$ in the presence of the
electromagnetic interaction.

\subsection{Effective potential}

For the BSW mechanism, an important problem is whether a test particle with
the critical angular momentum or critical charge can fall into the horizon.
We would like to investigate this problem from the viewpoint of effective
potential $V_{\text{eff}}=-\dot{r}^{2}/2$, where the dot denotes the
derivative with respect to the proper time. After doing the replacement (\ref%
{replacement}) and setting $m_{0}=1$ for simplicity, the effective potential
is%
\begin{equation}
V_{\text{eff}}=-\frac{g_{5}}{2g_{2}g_{6}}\left[ g_{2}\Xi
^{2}-(r-r_{H})^{n}(\Lambda ^{2}+g_{2})g_{6}\right] .  \label{Veff}
\end{equation}%
For a non-degenerated horizon, the effective potential for the particle with
critical angular momentum (\ref{Lc}) can be expanded near the horizon
\begin{equation}
V_{\text{eff}}(L_{c})|_{n=1}=\frac{g_{5}}{2}\left( 1+\frac{\Theta ^{2}}{%
g_{2}w^{2}}\right) (r-r_{H})+\mathcal{O}\left( r-r_{H}\right) ^{2}.
\label{n=1}
\end{equation}%
Since $g_{5}(r_{H})$ and $g_{2}(r_{H})$ are both positive for event
horizons, one has $V_{\text{eff}}>0$ which violates the positiveness of $%
\dot{r}^{2}$. So this particle can not fall into the event horizon. For a
usual two-fold degenerate horizon, the expansion of effective potential is
\begin{equation}
V_{\text{eff}}(L_{c})|_{n=2}=\frac{g_{5}}{2g_{2}g_{6}w^{2}}\left[
g_{6}(\Theta ^{2}+g_{2}w^{2})-g_{2}(qw\Phi +\Theta w^{\prime })^{2}\right]
(r-r_{H})^{2}+\mathcal{O}\left( r-r_{H}\right) ^{3},  \label{VeffL2}
\end{equation}%
where $\Phi =B^{\prime }+wC^{\prime }$ and the prime denotes the derivative
with respect to $r$. From this equation, we know that the particle with
critical angular momentum can exist in the region near a two-fold degenerate
horizon, if the coefficient of term$\sim (r-r_{H})^{2}$ in Eq. (\ref{VeffL2}%
) is negative. For the case of a multiple-fold degenerate horizon we can
prove by following steps that $V_{\text{eff}}|_{n\geq 3}$ has the same form.
First of all, we separate $V_{\text{eff}}$ into the terms with and without $%
(r-r_{H})^{n}$, which reads%
\begin{equation}
V_{\text{eff}}=\frac{(r-r_{H})^{n}\left( \Lambda ^{2}+g_{2}\right) g_{5}}{%
2g_{2}}-\frac{g_{5}\Xi }{2g_{6}}.  \label{prove}
\end{equation}%
Then, we substitute the value of $L_{c}$ into the second term and expand it
near the horizon as
\begin{equation}
-\frac{g_{5}\Xi (L_{c})}{2g_{6}}=-\frac{g_{5}(qw\Phi +\Theta w^{\prime
})^{2}(r-r_{H})^{2}}{2g_{6}w^{2}}+\mathcal{O}\left( r-r_{H}\right) ^{3}.
\label{prove2}
\end{equation}%
For $n\geq 3$ case, one can find that, comparing with Eq. (\ref{prove2}),
the first term of Eq. (\ref{prove}) is the higher order small quantity while
$r\rightarrow r_{H}$. So when $n\geq 3$, the expansion of effective
potential has the same form:%
\begin{equation}
V_{\text{eff}}(L_{c})|_{n\geq 3}=-\frac{g_{5}(qw\Phi +\Theta w^{\prime
})^{2}(r-r_{H})^{2}}{2g_{6}w^{2}}+\mathcal{O}\left( r-r_{H}\right) ^{3}.
\label{n>2}
\end{equation}%
Since the coefficient of term$\sim (r-r_{H})^{2}$ is always negative for
event horizons, we can conclude that the particle with $L_{c}$ can exist
near the region of a multiple-fold degenerate horizon without any
restriction that is needed for the case of two-fold degenerate horizons.

Now we are going to discuss whether a particle can touch the horizon from
infinity. It is usually judged by comparing the $L_{c}$ with the minimum and
maximum angular momentum ($L_{\min }$, $L_{\max }$) at circular orbits $r_{%
\text{cir}}$ solved from $V_{\text{eff}}=0$ and $\partial _{r}V_{\text{eff}%
}=0$. Obviously, we can not follow this approach directly since ($r_{\text{%
cir}}$, $L_{\min }$, $L_{\max }$) can not be solved in a general background.
However, it is interesting to see that we still can make the judgement in
general.

Setting $V_{\text{eff}}(r_{\text{cir}})=0$, we get the relationship between
the radial coordinate of these circular orbits and angular momentum:
\begin{eqnarray}
L(r_{\text{cir}}) &=&\frac{1}{g_{2}w^{2}-(r-r_{H})^{n}g_{6}}\bigg[%
(r-r_{H})^{n}qCg_{6}+g_{2}w(\Theta -qCw)  \nonumber \\
&&\pm \sqrt{(r-r_{H})^{n}g_{6}g_{2}\left[ \Theta
^{2}-(r-r_{H})^{n}g_{6}+g_{2}w^{2}\right] }\bigg]|_{r=r_{\text{cir}}}.
\nonumber
\end{eqnarray}%
From this equation, it is easy to notice that $L(r_{H})=L_{c}$, i.e., for a
test particle with critical angular momentum, it has a possible circular
orbit on the horizon. In order to ensure that this is a true circular orbit,
$\partial _{r}V_{\text{eff}}=0$ must also be satisfied. For a
non-degenerated horizon, $\partial _{r}V_{\text{eff}}$ on the horizon is%
\[
\partial _{r}V_{\text{eff}}(L_{c})|_{n=1}=\frac{g_{5}}{2}(1+\frac{\Theta ^{2}%
}{g_{2}w^{2}})|_{r=r_{H}}.
\]%
It is nonvanishing, which means that there is no circular orbit for a test
particle with critical angular momentum on the horizon. However, it is
obvious that $\partial _{r}V_{\text{eff}}=0$ on any-fold degenerate
horizons, since it is at least the second order function of $r-r_{H}$ as
just being seen in Eqs. (\ref{VeffL2}) and (\ref{n>2}). This indicates that $%
r_{H}$ is just the innermost circular orbit of the particle with $L_{c}$ for
any-fold degenerate horizons. Furthermore, because the existence of
particles near a degenerate horizon requires that the coefficients of term$%
\sim (r-r_{H})^{2}$ in Eqs. (\ref{VeffL2}) and (\ref{n>2}) are negative,
which means%
\begin{equation}
\partial _{r}^{2}V_{\text{eff}}(L_{c})|_{n\geq 2,r=r_{H}}<0,  \label{Vrr2}
\end{equation}%
the potential just has the maximum on the horizon. Hence the particle with $%
L_{c}$ can touch any-fold degenerate horizons from infinity (Rigorously, we
also need to assume that the potential is so ordinary that the maximum is
the global maximum, just as in the cases of Kerr and RN BHs where the
effective potential has only one maximum. Note that the maximum at the
horizon is obvious not the global maximum for the BH imbedded in AdS
background where the potential is divergent at infinite \cite{Jie Yang}.)
For example, Eq. (\ref{Vrr2}) is reduced to%
\[
\partial _{r}^{2}V_{\text{eff}}(L_{c})|_{n=2,r=r_{H}}=1-3E^{2}<0
\]%
for Kerr BHs, which can be satisfied for $E>\frac{1}{\sqrt{3}}$.

For charged and non-rotating BHs, $L_{c}$ and $1/w$ are infinite. So the
result gotten above is invalid and the above discussion process should be
repeated by replacing the critical angular momentum with the critical
charge. We briefly present the results as follows. Substituting Eq. (\ref{qc}%
) into Eq. (\ref{Veff}), the effective potentials for different $n$ are%
\begin{eqnarray}
V_{\text{eff}}(q_{c})|_{n=1} &=&\frac{g_{5}}{2}\left( 1+\frac{1}{g_{2}}\frac{%
\Psi ^{2}}{\Omega ^{2}}\right) (r-r_{H})+\mathcal{O}\left( r-r_{H}\right)
^{2},  \label{Veffq1} \\
V_{\text{eff}}(q_{c})|_{n=2} &=&\frac{g_{5}}{2g_{2}g_{6}}\left[ g_{6}\left(
\Psi ^{2}+g_{2}\Omega ^{2}\right) -g_{2}(\Gamma \Phi +\Psi w^{\prime })^{2}%
\right] (r-r_{H})^{2}+\mathcal{O}\left( r-r_{H}\right) ^{3},  \label{Veffq2}
\\
V_{\text{eff}}(q_{c})|_{n\geq 3} &=&-\frac{g_{5}}{2g_{6}\Omega ^{2}}(\Gamma
\Phi +\Psi w^{\prime })^{2}(r-r_{H})^{2}+\mathcal{O}\left( r-r_{H}\right)
^{3},  \label{Veffq3}
\end{eqnarray}%
where%
\[
\Psi _{i}=L_{i}B+E_{i}C,\;\Omega =B+Cw,\;\Gamma _{i}=E_{i}-L_{i}w.
\]%
One can conclude that the particle with critical charge can exist near
two-fold degenerate horizons, and can exist near multiple-fold degenerate
horizons. From Eqs. (\ref{Veffq1}), (\ref{Veffq2}) and (\ref{Veffq3}), one
can see that $r_{H}$ is just the innermost circular orbit of the particle
with $q_{c}$ for any-fold degenerate horizons but not for non-degenerate
horizons. If the following condition%
\[
\partial _{r}^{2}V_{\text{eff}}(q_{c})|_{n\geq 2,r=r_{H}}<0
\]%
is satisfied, the particle with $q_{c}$ can touch any-fold degenerate
horizons from infinity. For RN BHs as an instance, the condition is reduced
to%
\[
\partial _{r}^{2}V_{\text{eff}}(q_{c})|_{n=2,r=r_{H}}=1-E^{2}+L^{2}<0.
\]

\subsection{Multiple scattering}

Even for $n=1$ case where a test particle with $L_{c}$ can not fall into the
horizon, there is still a possible mechanism to achieve ultra-high CM
energy. Consider a particle with angular momentum close to the critical
value, i.e. $L=L_{c}(1-\delta )$, where $\delta $ is an arbitrary positive
number. At the horizon, its effective potential is%
\[
V_{\text{eff}}|_{r=r_{H}}=-\frac{\delta ^{2}g_{5}\left( L_{c}w\right) ^{2}}{%
2g_{6}}.
\]%
For $g_{5}(r_{H})/g_{6}(r_{H})>0$, the effective potential will be negative,
which means a test particle with angular momentum $L=L_{c}(1-\delta )$ can
exist in the region close to the event horizon. The value of this angular
momentum may be too large such that the test particle can not fall into the
horizon far from the horizon directly. However, it is possible that a test
particle with small angular momentum is ingoing to the horizon and interacts
with other particles on the accretion disc or decays to be more light
particles so that it gets larger angular momentum. This is the so-called
multiple scattering mechanism proposed in Ref. \cite{pavlov}. The
corresponding CM energy can be calculated as
\begin{equation}
\frac{E_{c.m.}^{2}}{2m_{0}^{2}}=1-\frac{\Theta _{1}\Theta _{2}}{g_{2}w^{2}}+%
\frac{\Xi _{2}\left( \Theta _{1}^{2}+g_{2}w^{2}\right) }{2g_{2}w^{3}L_{c}}%
\frac{1}{\delta }+\mathcal{O}\left( \delta \right) ^{1}.  \label{CMM1}
\end{equation}%
On the other hand, the charge of the particle can be amplified to $%
q=q_{c}(1-\delta )$ by the multiple scattering mechanism, since the pairs of
electron and positron could be created by the collision of the high energy
photons and massive atoms. For this particle, its effective potential is%
\[
V_{\text{eff}}|_{r=r_{H}}=-\frac{\delta ^{2}g_{5}(E-Lw)^{2}}{2g_{6}}.
\]%
Similar to the former case, this test particle can exist in the region close
to the horizon, and the CM energy is
\begin{equation}
\frac{E_{c.m.}^{2}}{2m_{0}^{2}}=1-\frac{\Psi _{1}\Psi _{2}}{g_{2}\Omega ^{2}}%
+\frac{\Xi _{2}\left( \Psi _{1}^{2}+g_{2}\Omega ^{2}\right) }{2g_{2}\Gamma
_{1}\Omega ^{2}}\frac{1}{\delta }+\mathcal{O}\left( \delta \right) ^{1}.
\label{CMM2}
\end{equation}%
When $\delta \rightarrow 0$, both CM energy (\ref{CMM1}) and (\ref{CMM2})
will be arbitrary high.

\subsection{Proper time problem}

Now, we would like to have a glance on the proper time required for a test
particle to reach the horizon. It can be obtained from%
\begin{equation}
\tau =\int_{r_{i}}^{r_{f}}\left( \frac{1}{\sqrt{-2V_{\text{eff}}(r)}}\right)
dr.  \label{TT}
\end{equation}%
Since the effective potentials are the first order functions of $(r-r_{H})$
in Eqs. (\ref{n=1}) and (\ref{Veffq1}), the proper time for the particle
with critical angular momentum or charge falling into the non-degenerate
horizon will be finite. On the contrary, a test particle with critical
angular momentum or charge takes infinite proper time to fall into any-fold
degenerate horizons, because the effective potentials (\ref{VeffL2}), (\ref%
{n>2}), (\ref{Veffq2}), and (\ref{Veffq3}) are the second order functions of
$(r-r_{H})$, which means that the proper time is logarithmic divergent.

\subsection{ISCO for the test particle with critical angular momentum or
charge}

To obtain an arbitrary high CM energy, the angular momentum and charge of a
test particle must be fine-tuned. The existence of ISCO has been realized as
a possibility to solve this problem \cite{ISCO}. Now we will extend the
discussion of the ISCO in Kerr BHs \cite{ISCO} to a general case. We point
out that the key point to get arbitrary high CM energy without the
fine-tuning problem is to require the ISCO with critical angular momentum or
charge exactly located on the horizon. In other words, the CM energy of one
particle collided with another one moving along the ISCO is finite, unless
the ISCO is located on the horizon and the particle along the ISCO has the
critical parameters. For a two-fold degenerate horizon, by setting $\partial
_{r}^{2}V_{\text{eff}}=0$ where $V_{\text{eff}}$ is given by (\ref{VeffL2}),
we obtain the condition to require the ISCO with critical angular momentum
just on the horizon, which reads%
\[
g_{6}(\Theta ^{2}+g_{2}w^{2})-g_{2}(qw\Phi +\Theta w^{\prime
})^{2}|_{r=r_{H}}=0.
\]%
For $n\geq 3$ cases, from Eq. (\ref{n>2}), we know that the condition for
ISCO on the horizon is%
\[
qw\Phi +\Theta w^{\prime }|_{r=r_{H}}=0,
\]%
which can be rewritten clearly as
\begin{equation}
E=-\frac{q(wB^{\prime }+w^{2}C^{\prime }-Bw^{\prime })}{w^{\prime }}%
|_{r=r_{H}}.  \label{electro}
\end{equation}%
The case of a test particle with critical charge has similar results. We can
obtain%
\[
g_{6}\left( \Psi ^{2}+g_{2}\Omega ^{2}\right) -g_{2}(\Gamma \Phi +\Psi
w^{\prime })^{2}|_{r=r_{H}}=0
\]%
for $n=2$, and%
\[
\Gamma \Phi +\Psi w^{\prime }=0
\]%
for $n\geq 3$ cases, which can be recast as%
\begin{equation}
E=\frac{L\left( wB^{\prime }+w^{2}C-Bw^{\prime }\right) }{B^{\prime
}+wC^{\prime }+Cw^{\prime }}|_{r=r_{H}}.  \label{frame}
\end{equation}%
From Eqs. (\ref{electro}) and (\ref{frame}), it is interesting to note that
both frame dragging effect and electromagnetic interaction are necessary for
ISCO on the multiple-fold degenerate horizons, provided that the energy is
nonvanishing.

\subsection{Duality between frame dragging effect and electromagnetic
interaction}

In Ref. \cite{shell}, an upper limit was found to exist for the total energy
of colliding shells in the observable domain in the BSW process due to the
gravity of the shells. Although this result is obtained in the RN
background, since RN BHs are easily to be tackled based on their higher
symmetry than Kerr BHs, it has been suspected that an upper limit might also
exist for the Kerr background, noticing the similarity of BSW mechanism in
Kerr and RN BHs.

Here we would like to clarify the corresponding relationship between the
frame dragging effect and the electromagnetic interaction in the BSW
mechanism from the viewpoint of critical angular momentum and charge,
effective potential and CM energy, which are three essential factors in the
BSW mechanism.

In Eq. (\ref{qc}) with $w(r)=0$, by doing the transformation%
\begin{equation}
B(r)\rightarrow w(r)\quad \text{and}\quad q\rightarrow L,
\label{transformation}
\end{equation}%
Eq. (\ref{qc}) will equate to Eq. (\ref{Lc}) with $q=0$. Therefore, from the
viewpoint of critical angular momentum and critical charge, there exists an
exact duality. Then, we are interested in the duality of effective
potential. We can expand the effective potential for a rotating non-charged
BH background by setting $q=0$, $B(r)=0$ and $C(r)=0$, which reads%
\begin{equation}
\frac{g_{3}(r)}{2}-\frac{E^{2}g_{3}(r)}{2g_{1}(r)}+\frac{L^{2}g_{3}(r)}{%
2g_{2}(r)}+\frac{ELg_{3}(r)w(r)}{g_{1}(r)}-\frac{L^{2}g_{3}(r)w^{2}(r)}{%
2g_{1}(r)}.  \label{rotating EV}
\end{equation}%
We also expand it for a static charged BH background by setting $L=0$ and $%
w(r)=0$,\ which gives%
\begin{equation}
\frac{g_{3}(r)}{2}-\frac{E^{2}g_{3}(r)}{2g_{1}(r)}+\frac{qEB(r)g_{3}(r)}{%
g_{1}(r)}-\frac{q^{2}B(r)^{2}g_{3}(r)}{2g_{1}(r)}.  \label{charged EV}
\end{equation}%
After doing the transformation (\ref{transformation}), one can find that Eq.
(\ref{charged EV}) is the same as Eq. (\ref{rotating EV}), up to only one
term $\frac{L^{2}g_{3}(r)}{2g_{2}(r)}$. We also notice that this term will
be vanished on the horizon, where $g_{3}(r)\rightarrow 0$. The last step is
to consider the CM energy. Eq. (\ref{CMn}) for static charged BH background
is%
\begin{equation}
\frac{E_{c.m.}^{2}}{2m_{0}^{2}}=1+\frac{\Theta _{2}(r_{H})}{2\Theta
_{1}(r_{H})}+\frac{\Theta _{1}(r_{H})}{2\Theta _{2}(r_{H})}.
\label{charged CM}
\end{equation}%
After doing the transformation (\ref{transformation}), Eq. (\ref{CMn}) for
the rotating non-charged BH can be written as
\begin{equation}
\frac{E_{c.m.}^{2}}{2m_{0}^{2}}=1+\frac{\Theta _{2}(r_{H})}{2\Theta
_{1}(r_{H})}+\frac{\Theta _{1}(r_{H})}{2\Theta _{2}(r_{H})}+\frac{%
(E_{2}q_{1}-E_{1}q_{2})^{2}}{2\Theta _{1}(r_{H})\Theta
_{2}(r_{H})g_{2}(r_{H})}.  \label{rotating CM}
\end{equation}%
One can find that the difference between Eqs. (\ref{charged CM}) and (\ref%
{rotating CM}) is the last term of (\ref{rotating CM}). Since the charge
energy ratio of two particles should not be the same for gaining ultra-high
CM energy, this term is nonvanishing, and it will diverge as well as all
other terms in (\ref{rotating CM}) when one of the colliding particle has
the critical charge. Thus, we can conclude that the duality between frame
dragging effect and electromagnetic interaction is not exact, but if one is
only interested in the properties on the horizon, this duality is
qualitatively effective.

\section{Braneworld black holes}

In this section, we will apply some obtained general results to braneworld
BHs. It is very interesting since braneworld BHs not only could exist as
astrophysical BHs, but also could be produced at LHC, which hence provides a
possibility to check the BSW mechanism terrestrially.

In braneworld theory, the standard model particles are confined on the
brane, with only gravity propagating in the bulk. We will assume that the
test particle moves on the equatorial plane of braneworld BHs.

\subsection{ADD KNM BHs}

The ADD model has $d$ flat, compact extra dimensions. Assuming the 3-brane
located at $\theta _{i}$ $=\pi /2$ ($i=1,\cdots ,d$), the metric of the
equatorial plane of KNM BHs in this model is \cite{Kanti}
\begin{equation}
ds^{2}=-\frac{\Delta -a^{2}}{\Sigma }dt^{2}-\frac{2a(a^{2}+r^{2}-\Delta )^{2}%
}{\Sigma }dtd\phi +\frac{(a^{2}+r^{2})^{2}-a^{2}\Delta }{\Sigma }d\phi ^{2}+%
\frac{r^{2}}{\Delta }dr^{2},  \label{ADD metric}
\end{equation}%
where
\[
\Delta =r^{2}+a^{2}+Q^{2}-\frac{\mu }{r^{d-1}},\qquad \Sigma =r^{2},
\]%
$Q$ is the charge of the BH, $\mu $ is the mass constant and will be set as $%
2$ for convention. Comparing this metric with Eq. (\ref{metric}), we obtain
the expression of those functions in Eq. (\ref{metric}):
\begin{equation}
g_{1}(r)=\frac{\Delta \Sigma }{(a^{2}+r^{2})^{2}-\Delta a^{2}},\quad
g_{2}(r)=\frac{(a^{2}+r^{2})^{2}-\Delta a^{2}}{\Sigma },  \label{AKN1}
\end{equation}%
\begin{equation}
g_{3}(r)=\frac{\Delta }{\Sigma },\quad w(r)=\frac{a(a^{2}+r^{2}-\Delta )}{%
(a^{2}+r^{2})^{2}-\Delta a^{2}}.  \label{AKN2}
\end{equation}%
Correspondingly, the electromagnetic potential is
\[
A_{t}=B(r)=-\frac{Qr}{\Sigma },\qquad A_{\phi }=C(r)=\frac{Qar}{\Sigma }.
\]%
When $d=0$, the metric (\ref{ADD metric}) reduces to the KNM metric in
four-dimensional spacetime. To investigate the property of the horizon of
ADD KNM BHs, let us set $\partial _{r}\Delta =0$, this leads to
\[
r=(1-d)^{\frac{1}{1+d}}.
\]%
This equation is important since it indicates that there is no degenerated
horizon for ADD KNM BHs, when $d>0$.

Now, let us use the six-dimensional ADD KNM BH as an example to show the BSW
mechanism in the ADD model (since\ the five-dimensional ADD model has been
ruled out). We assume the angular momentum $a\leq 0.998$ and charge $Q\leq
\frac{4}{3}\sqrt{1/137}=0.113$ \cite{Sampaio}. For a six-dimensional charged
ADD BH with $a=0.998$ and $Q=0.100(<0.113)$, the location of horizon becomes
to%
\[
r_{H}=\frac{-2^{1/3}(a^{2}+Q^{2})}{\left[ 54+\sqrt{2916+108(a^{2}+Q^{2})^{3}}%
\right] ^{1/3}}+\frac{\left[ 54+\sqrt{2916+108(a^{2}+Q^{2})^{3}}\right]
^{1/3}}{3\times 2^{1/3}}=0.999,
\]%
and the critical angular momentum (\ref{Lc}) and charge (\ref{qc}) are $%
L_{c}=2.00E-0.100q$ and $q_{c}=20.0E-10.0L$, respectively.

We note that it is difficult to solve $V_{\text{eff}}=0\quad $and$\quad
\partial _{r}V_{\text{eff}}=0$ analytically to determine whether the
particle with $L_{c}$ or $q_{c}$ can fall into the BH. However, from Eqs. (%
\ref{n=1}) and (\ref{Veffq1}), and $g_{5}(r_{H})=4.02>0$, one immediately
recognizes that the answer is negative.

The multiple scattering process is necessary to achieve ultra-high CM energy
in this case. For the particle with $L_{1}=L_{c}(1-\delta )$ or $%
q_{1}=q_{c}(1-\delta )$ colliding with another particle on the horizon, the
CM energy (\ref{CMM1}) and (\ref{CMM2}) are
\begin{equation}
\frac{E_{c.m.}^{2}}{2m_{0}^{2}}=1+\frac{0.501\left(
0.999+E_{1}^{2}+0.200E_{1}q_{1}+0.010q_{1}^{2}\right)
(E_{2}-0.501L_{2}+0.050q_{2})}{E_{1}-0.050q_{1}}\frac{1}{\delta }+\mathcal{O}%
\left( \delta \right)  \label{AE1}
\end{equation}%
and%
\begin{equation}
\frac{E_{c.m.}^{2}}{2m_{0}^{2}}=1+\frac{50.0\left(
0.010+0.010E_{1}^{2}-0.020E_{1}L_{1}+0.010L_{1}^{2}\right)
(E_{2}-0.501L_{2}+0.050q_{2})}{E_{1}-0.501L_{1}}\frac{1}{\delta }+\mathcal{O}%
\left( \delta \right) .  \label{AE2}
\end{equation}%
We note that $E_{1}-0.050q_{1}\sim L_{c}$ and $E_{1}-0.501L_{1}\sim q_{c}$
can not be vanishing, and the divergent degree (index of $\delta $) is not
influenced by $q_{1}$ in Eq. (\ref{AE1}) and $L_{1}$ in Eq. (\ref{AE2}).

\subsection{RS tidal charge BHs}

After having investiating the ADD KNM BH in which the horizon is
non-degenerated, we would like to study the degenerated horizon of an
extreme RS tidal charged BH.

The RS model consists of a single, positive tension brane in an infinite
extra dimension. The role of extra dimension is played by a tidal charge $Q$%
. The effective metric of a RS BH can be expressed like \cite{braneworld BH}
\begin{equation}
ds^{2}=-\left( 1-\frac{2Mr-Q}{\Sigma }\right) dt^{2}-2\frac{a(2Mr-Q)}{\Sigma
}dtd\phi +\frac{\Sigma }{\Delta }dr^{2}+\left( r^{2}+a^{2}+\frac{2Mr-Q}{%
\Sigma }a^{2}\right) d\phi ^{2},  \nonumber
\end{equation}%
where
\[
\Delta =r^{2}-2Mr+a^{2}+Q,\;\Sigma =r^{2}.
\]%
Note that the mass $M$\ of BHs will be set as $1$\ later. By comparing with
Eq. (\ref{metric}), one can find
\[
g_{1}(r)=\frac{\Delta \Sigma }{(a^{2}+r^{2})^{2}-\Delta a^{2}},\quad
g_{2}(r)=\frac{(a^{2}+r^{2})^{2}-\Delta a^{2}}{\Sigma },
\]%
\[
g_{3}(r)=\frac{\Delta }{\Sigma },\quad w(r)=\frac{a(a^{2}+r^{2}-\Delta )}{%
(a^{2}+r^{2})^{2}-\Delta a^{2}}.
\]%
If $a^{2}+Q=1$, this BH has a two-fold degenerate horizon located at $%
r_{H}=1 $. Under the limit of the solar system on the tidal charge $Q\leq
8\times 10^{4}$m$^{2}=0.037$\ for a BH with one sun mass \cite{hmer}, the
horizon can be degenerated if $a\geqslant 0.981$. From the discussion about
two-fold degenerated horizons in the preceding section, we know that, for
the extreme RS tidal charge BH with $a=\sqrt{1-Q}$, the CM energy of two
particles will be divergent, provided that one particle's angular momentum
equates to the critical value (\ref{Lc})%
\begin{equation}
L=E\frac{2-Q}{\sqrt{1-Q}},  \label{LcR}
\end{equation}%
and the condition (\ref{Vrr2})%
\[
1+(\frac{1}{1-Q}-4)E^{2}<0
\]%
is satisfied. This inequality is saturated when%
\begin{equation}
E=\sqrt{\frac{1-Q}{3-4Q},}  \label{ER}
\end{equation}%
which is one of the conditions for ISCO on the horizon that can be invoked
to avoid the fine-turning problem.

An important problem of the BSW mechanism in braneworld BHs is about the
infinite proper time spent by a particle with $L_{c}$ on falling into the
degenerated horizon of BH. One may worry about whether the ultra-high energy
can be achieved rapidly in small braneworld BHs, since in four-dimension
case, the lifetime of BH decreases with its mass $M$ rapidly: $\tau \approx
8.3\times 10^{-26}\frac{M^{3}}{1g}s$ \cite{Hartle}. Considering the
particles colliding at $r_{f}=1+\delta ^{\prime }$ near the horizon, and
expanding Eq. (\ref{uc1}) near the horizon $r_{H}=1$, we have%
\begin{equation}
\frac{E_{c.m.}^{2}}{2m_{0}^{2}}=1+\frac{\alpha }{\delta ^{\prime }}+\mathcal{%
O}(\delta ^{\prime })^{0},  \label{Ecm1}
\end{equation}%
where%
\[
\alpha =\left( E_{2}\frac{2-Q}{\sqrt{1-Q}}-L_{2}\right) \left( 2E_{1}-\sqrt{%
E_{1}^{2}\frac{3-4Q}{1-Q}-1}\right) .
\]%
From Eq. (\ref{TT}), one can obtain the proper time cost for achieving the
energy expressed in (\ref{Ecm1}) as%
\begin{eqnarray}
\tau &\approx &-\frac{\sqrt{1-Q}}{\sqrt{\left( 4E_{1}^{2}-1\right) \left(
1-Q\right) -E_{1}^{2}}}\log \delta ^{\prime }  \nonumber \\
&=&\frac{\sqrt{1-Q}}{\sqrt{E_{1}^{2}(3-4Q)+Q-1}}\left[ \log \left( \frac{%
E_{c.m.}^{2}}{2m_{0}^{2}}-1\right) -\log \alpha \right] ,  \label{T}
\end{eqnarray}%
where we have omitted the effect from a finite $r_{i}$. As an example, we
will study the proper time spent by a particle on falling into to a micro RS
BH with mass $M$. For this BH, we would like to consider how long the BSW
process would take to get $1$TeV energy, which is a significant scale
denoting both the quantum gravity scale in extra-dimensional theory and the
mass of BHs. From Eq. (\ref{T}), we obtain $\tau \approx 10^{-59}\frac{M}{1%
\text{TeV}}s$, assuming $m_{0}$ as the electron mass (noting that other
parameters are not important). This value is larger than the lifetime of
four dimensional TeV BHs $\sim 10^{-88}s$, but it is smaller than the
typical lifetime of small braneworld BHs, which is about $10^{-26}s$ for ADD
BHs \cite{Kanti}, and can even reach up to $10^{9}s$ for RS cases \cite{RST}.

One can solve the ISCO for a RS BH from the equations
\begin{equation}
V_{\text{eff}}=0,\quad \partial _{r}V_{\text{eff}}=0\quad \text{and}\quad
\partial _{r}^{2}V_{\text{eff}}=0.  \label{ISCO1}
\end{equation}%
Setting $a=1$ and $Q=0$, the particle on the ISCO satisfies%
\[
L=\frac{2}{\sqrt{3}},\qquad E=\frac{1}{\sqrt{3}},\qquad r=1,
\]%
consistent with the result for Kerr BHs. For $Q\neq 0$ case, Eq. (\ref{ISCO1}%
) is difficult to solve directly. However, based on the fact that the ISCO
should be on the horizon, we have obtained the parameters of ISCO, which are
given by Eqs. (\ref{LcR}) and (\ref{ER}). One can substitute the value of $L$
and $E$ into $\partial _{r}^{2}V_{\text{eff}}$ and check that it is zero
indeed. It is also easy to notice that Eqs. (\ref{LcR}) and (\ref{ER}) will
go back to the ISCO of Kerr BHs with $Q=0$. To solve the fine-tuning
problem, one should consider ISCO on $r=1+\delta ^{\prime }$ near the
horizon of the RS BH with $a=\sqrt{1-Q}(1-\delta )$. The event horizon of
this BH locates at $r_{H}=1+\sqrt{(1-Q)(2-\delta )\delta }$. We can also
solve the $E(r)$ and $L(r)$ of circular orbits for this case. Then we will
solve $\partial _{r}E(r)=0$ and $\partial _{r}L(r)=0$, instead of
calculating $\partial _{r}^{2}V_{\text{eff}}=0$ directly, which leads to
\begin{equation}
\delta =\frac{(1-2Q)\delta ^{\prime 3}}{4(1-Q)^{2}}+\mathcal{O}(\delta
^{\prime })^{4}.  \label{ddp}
\end{equation}%
Note that one must impose $1-2Q\gg 0$ to preserve the effective
approximation in Eq. (\ref{ddp}). Thus, we get the expression of $E$ and $L$
for the ISCO on $r=1+\delta ^{\prime }$, which reads
\begin{eqnarray}
E &=&\sqrt{\frac{1-Q}{3-4Q}}+\frac{3(1-2Q)\left( 1-Q\right) ^{1/2}}{\left(
3-4Q\right) ^{3/2}}\delta ^{\prime }+\mathcal{O}(\delta ^{\prime })^{2},
\label{ISCO E} \\
L &=&\frac{2-Q}{\sqrt{3-4Q}}+\frac{3\left( 2-5Q+2Q^{2}\right) }{\left(
3-4Q\right) ^{3/2}}\delta ^{\prime }+\mathcal{O}(\delta ^{\prime })^{2}.
\label{ISCO L}
\end{eqnarray}%
Obviously, the value of $E$ and $L$ is the same as Eqs. (\ref{LcR}) and (\ref%
{ER}) when $\delta ^{\prime }=0$. We consider the collision with one
particle on this ISCO and another particle with $E=1$ and $L=0$ for
simplicity. The CM energy is
\begin{equation}
\frac{E_{c.m.}^{2}}{2m_{0}^{2}}=1+\frac{\sqrt{2}(2-Q)\left( 1-Q\right) }{%
\sqrt{\left( 3-4Q\right) \left( 1-2Q\right) }}\frac{1}{\delta ^{\prime 3/2}}-%
\frac{3(1-2Q)^{3/2}(2-Q)}{\sqrt{2}(3-4Q)^{3/2}}\frac{1}{\delta ^{\prime 1/2}}%
+\mathcal{O}\left( \delta ^{\prime }\right) ^{1/2}.  \label{ISCO CM E 2}
\end{equation}%
We notice that the CM energy can be arbitrary high when the value of $\delta
^{\prime }$ is arbitrary small. Also, the divergent degree is not influenced
by the tidal charge.

\section{Conclusion and discussion}

BSW mechanism provides a remarkable possibility that Kerr BHs might act as
Plank-scale particle accelerators. Some extended works showed that the
frame-dragging effect is important to achieve unbound CM energy. It was
further confirmed in \cite{general rotating}, where the arbitrary high CM
energy was found as a general property in a general axially symmetric
rotating BH. After that, however, there were still some works on complicated
rotating backgrounds, partially because the metric adapted in \cite{general
rotating} and the consequent results, such as the expression of CM\ energy,
can not be compared with those concrete backgrounds directly. Moreover, the
ISCO in Kerr BHs, which was introduced to avoid the fine-turning problem
\cite{ISCO}, has not been applied to a general background. On the other
hand, it was found that the collision of charged particles in RN BHs has the
similar mechanism for arbitrary high CM energy, but the general charged
background has not been considered either. In particular, the combined
effect of frame dragging and electromagnetic interaction is very worth to be
studied, because theoretically, it could provide a better understanding of
these two kinds of effects in BSW mechanism; and practically, the ADD KNM BH
is a general BH that could be formed after proton-proton collisions in LHC
\cite{Sampaio}.

Our work addressed these problems mentioned above. We investigated the CM
energy of two charged particles colliding in the background of a general
stationary charged BH, adapting a metric which is convenient to compare with
observations. It is shown that the CM energy can be arbitrarily high,
provided that the following three conditions are satisfied. First, the
collision should occur near the horizon of BHs. Second, only one particle
has the entangled (near) critical angular momentum $L_{c}$ and critical
charge $q_{c}$. The last condition depends on the degenerate degree of
horizons. Concretely, for two-fold degenerate horizons, there are some
restrictions on the parameters of particles from the requirement that the
effective potential with $L_{c}$ or $q_{c}$ should be negative near
horizons. Since the negative potential near horizons just imposes the
maximum of potential at horizons, the particle with $L_{c}$ or $q_{c}$ under
the restrictions can reach the horizons even from infinity. For
multiple-fold degenerate event horizons, there is no such restriction. The
ultra-high CM energy can also be gained for the particle collision near
non-degenerate horizons by invoking the multiple scattering mechanism to
amplify the angular momentum or charge of the falling particle from
infinity. We derived the general formula of CM energy for non-degenerate and
any-fold degenerate horizons when one particle has $L_{c}$ or $q_{c}$.

Furthermore, we obtained the condition for the existence of ISCO with $L_{c}$
or $q_{c}$ on degenerate horizons, and pointed out that it is essential to
get arbitrary high CM energy without the fine-tuning problem. It is
interesting to see that both frame dragging effect and electromagnetic
interaction are necessary for the existence of ISCO on multiple-fold
degenerate horizons with nonvanishing energy parameter $E$. Moreover, we
showed that the proper time taken for achieving infinite CM energy is finite
for the particle collision at non-degenerate horizons but is logarithmic
divergent for the collision at degenerate horizons. We also clarified that
there is a qualitatively effective duality between frame dragging effect and
electromagnetic interaction for the properties of the BSW mechanism on the
horizon, which could be helpful to investigate whether the CM\ energy of
colliding particles around rotating BHs has a similar upper limit which was
found by studying the acceleration of colliding shells around a RN black
hole \cite{shell}.

It should be pointed out that we have ignored the back-reaction and
gravitational radiation of the colliding particles, which may have important
effect on the CM energy but could not be analyzed in the present general
frame.

We then applied some general results to the cases of ADD and RS braneworld
BHs. It was shown that there is no degenerate horizon in ADD KNM BHs, so we
calculated the CM energy with near critical angular momentum or near
critical charge obtained by the multiple scattering. It was found that the
divergent degree is not influenced by the charge or angular momentum of the
particle. For RS BHs, we found that the proper time spending by the
particles to arrive at 1TeV CM energy is smaller than the typical lifetime
of braneworld BHs. We also evaluated the CM\ energy of one particle
colliding with another particle on the ISCO when the horizon is near
degenerate. Also, the divergent degree is not influenced by the tidal charge.

At last, we expect that the BSW mechanism could be checked in LHC. We have
noted that the lifetime of these small BHs is long enough to afford the BSW
process to get the ultra-high energy. Thus, we can take the small BHs, which
are assumed to have been produced by proton-proton collisions in LHC, as the
background when the other two particles fall into and collide near the small
BHs. In other words, the braneworld small BHs in LHC plays the role of the
astrophysical black holes in the BSW mechanism. Although the small BH could
be distorted more easily than astrophysical BHs under the back reaction of
falling particles and the ultra-high CM energy could not be attained, one
still can expect that the CM energy of particles collided in the background
with small BHs would be apparently larger than the background without BHs.
Consider the ingoing particles being static at infinite and the mass of
small RS BHs is about $1$TeV. We can estimate the maximized CM energy after
counting the back-reaction effect, which can be implement by considering the
absorption of the first pair of colliding particles \cite{Comment}. From Eq.
(\ref{ISCO CM E 2}) and $\delta ^{\prime }\sim \left( m_{0}/M\right) ^{1/3}$%
, we have $E_{c.m.}\lesssim 6.5\left( \frac{m_{0}}{1\text{GeV}}\right)
^{3/4}\left( \frac{M}{1\text{TeV}}\right) ^{1/4}$GeV. This result means that
two particles colliding in an LHC experiment can reach higher energy due to
interaction with the small BHs, which have been produced by proton-proton
collisions before the BSW collision occurs. In particular, the CM energy
would be maximized when the angular momentum or charge of one particle
approaches the critical parameter $L_{c}$ or $q_{c}$, and decrease when
another particle also has the critical parameters. We expect that LHC could
check these unique properties of BSW mechanism. Nevertheless, it should be
noticed that our estimation is very rough since it has neglected the
underlying details of experiments.

\begin{acknowledgments}
SFW was partially supported by National Natural Science Foundation of China
(No. 10905037). YZ, SFW and YJ were also partially supported by Shanghai
Leading Academic Discipline Project No. S30105 and the Shanghai Research
Foundation No. 07dz22020. YXL was partially supported by the Program for New
Century Excellent Talents in University and the National Natural Science
Foundation of China (No. 10705013).
\end{acknowledgments}

\end{document}